\documentclass[doublecol]{epl2} 
\usepackage{amsmath}

\usepackage{amssymb}

\title{Wave transport in one-dimensional disordered systems with finite-width potential steps}
\shorttitle{Wave transport in 1-D disordered systems with finite-width steps} %Insert here a short version of the title if it exceeds 70 characters

\author{M. D\'iaz\inst{1} \and P. A. Mello\inst{1} \and M. Y\'epez\inst{2} \and S. Tomsovic\inst{3}}
\shortauthor{M. D\'iaz \etal}

\institute{                    
  \inst{1} Instituto de F\'{\i}sica, Universidad Nacional Aut\'{o}noma de M\'{e}xico - Apartado Postal 20-364, M\'{e}xico, D.F.   \\
  \inst{2} Departamento de F\'isica de la Materia Condensada,
Universidad Aut\'onoma de Madrid -  E-28049, Madrid, Spain  \\
  \inst{3} Department of Physics and Astronomy,
Washington State University - Pullman, WA}


\abstract{
An amazingly simple model of correlated disorder is a one-dimensional chain of $n$ potential steps with a fixed width $l_c$
and random heights.  
A theoretical analysis of the average transmission coefficient and Landauer resistance as functions of $n$ and $\delta=k l_c$ predicts two distinct regimes of behavior, one marked by extreme sensitivity and the other associated with exponential behavior of the resistance.  
The sensitivity arises in $n$ and $\delta$ for $\delta\approx \pi$, where the system is nearly transparent.  Numerical simulations match the predictions well, and they suggest a strong motivation for experimental study.
}

\begin{document}

\maketitle
%%%%%%%%%%%%%%%%%%%%%%%%%%%%%%%%%%%%%%%

%%%%%%%%%%%%%%%%%%%%%%%%%%%%%%%%%%%%%%%%%%%
\section{Introduction}

An enormous amount is known about wave transport in disordered systems described by a potential with uncorrelated disorder 
(see, e.g., Refs.~\cite{lifshitz_et_al,mello-kumar}
and references therein).  
The extension to studies of correlated disorder originated a couple of decades ago, when a number of surprising properties were found in one-dimensional (1D) systems.  Correlated disorder studies have since multiplied considerably, and cover a broad range of topics.  
For example, in the random-dimer model 
--a paradigm for the description of short-range correlations in 1D disordered systems-- fully transparent (delocalized) states have been discovered~\cite{dunlap_prl90,phillips91,bovier92,flores93}.  
For long-range correlated disorder, a mobility edge in 1D random potentials was found~\cite{izrailev-krochin99}, and the authors construct potentials with mobility edges at specific energies;  
one can thus build systems having the desired transport properties by controlling the correlations.  
Perhaps more surprising, an Anderson-like metal-insulator transition~\cite{moura98} and violations of single-parameter scaling due to short-range correlations~\cite{titov05} have been found.  
Ref. \cite{titov05} developed Fokker-Planck equations similar to the ones found in the present paper.  
Delocalization in the continuous random-dimer model~\cite{sanchez_izrailev_et_al},
and in continuous disordered systems consisting of $\delta$-potentials and 
barrier-well sequences \cite{lifshitz_et_al,hilke_flores} have also been discussed.

Here, wave transport in 1D disordered systems consisting of spatially extended scatterers --barriers and wells with a finite width-- is considered, in contrast to previous studies by our group~\cite{froufe_et_al_2007} in which the spatial extension of the scatterers and their separation were negligible and played no role in the analysis.  
The system contains $n$ steps, assumed to be low with respect
%%%%%%%%%%%%%%%%%%%%%%%%%%%%%%%%%%%%%%
\begin{figure}[ht]
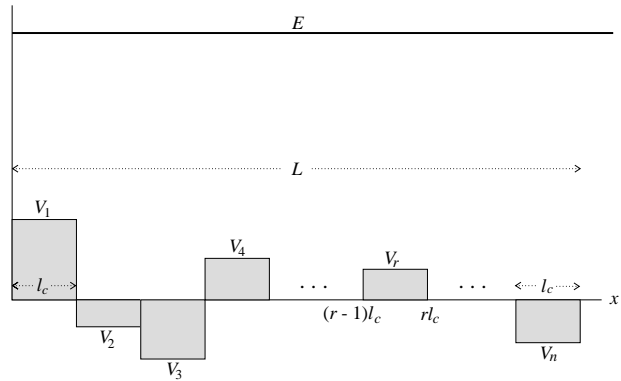

%\epsfxsize=0.40\textwidth
%\epsfysize=0.40\textwidth  
%\centerline{\epsffile{figure1version9.eps}}
\onefigure[scale=0.33]{Fig1.eps}
\caption{
\footnotesize{
Schematic representation of an array of $n$ steps of random height $V_r$ 
($r=1,\cdots n$) possessing fixed spatial width $l_c$.  
The incident energy $E$ is taken larger than all the $|V_r|$ 's.
}}
\label{rand_steps}
\end{figure}
%%%%%%%%%%%%%%%%%%%%%%%%%%%%%%%%%%%%%%
to the energy $E$ (see Fig. \ref{rand_steps}) and characterized by:  
i) a fixed width $l_c$ which may fit an arbitrary number of wavelengths 
$\delta/2 \pi$, where $\delta = k l_c$, with $k$ the wave number, and 
ii) random heights $V_r$ ($r=1,\cdots n$). 
The $n$ heights $V_r$ are statistically independent of one another, and identically and uniformly distributed, with zero average.  
In this model the correlations could scarcely be simpler, in the sense that for each step the potential is perfectly correlated within $l_c$, but perfectly uncorrelated otherwise. 
One motivation for studying this model is that it could be interpreted
(see Fig.~\ref{rand_steps}) as simulating a potential described by a random process with a correlation length $l_c$.
Another motivation is that it exhibits intriguing transport properties which we now describe.

%%%%%%%%%%%%%%%%%%%%%%%%%%%%%%%%%%%%
\subsection{Results and discussion}
\begin{figure}[ht]
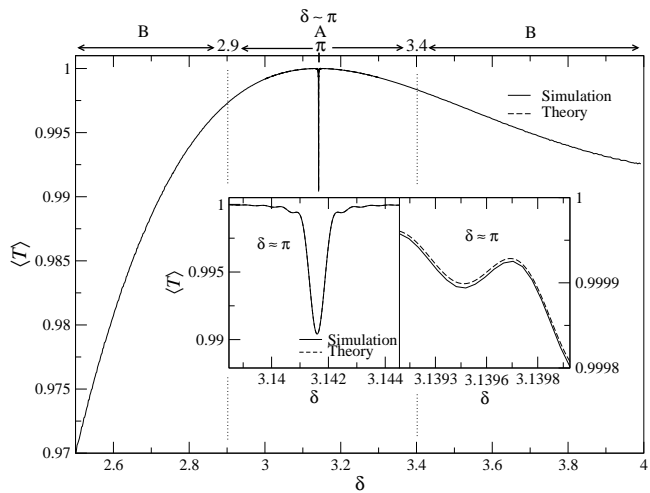

\onefigure[scale=0.35]{Fig2.eps}
\caption{
\footnotesize{
Theory and numerical simulations for $\langle T \rangle$ vs $\delta$, for a chain of $n=5000$ scatterers, $10^5$ realizations and for $y_0=0.09$ (the parameter related to the average step strength: see below Eq. (\ref{A(n)})).
The main figure shows the gross-structure behavior
as well as regimes A and B described in the text.  
The insets show the fine structure for $\delta \approx \pi$ and its enhanced behavior exactly at $\delta = \pi$, leading to a dip at this point.  This is consistent with the peak observed in Fig. \ref{av_resist_vs_delta} for the average resistance.  The agreement between simulation and theory is excellent.  The statistical error bar for $\delta=\pi$ is $\sim 10^{-5}$.
}
}
\label{avT_vs_delta}
\end{figure}

\begin{figure}[ht]
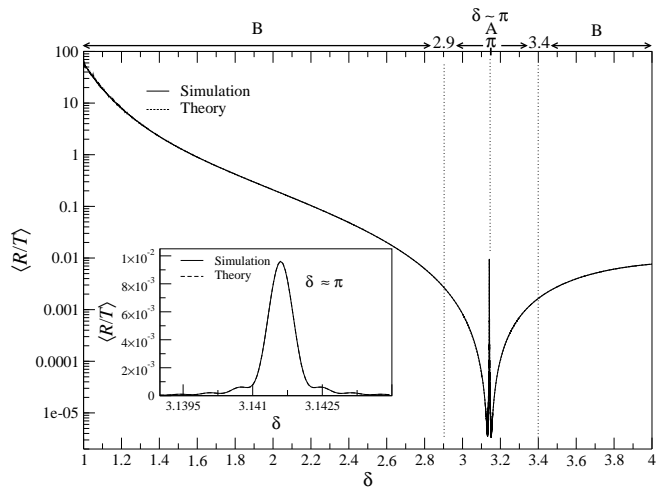

\onefigure[scale=0.35]{Fig3.eps}
\caption{
\footnotesize{
Theory and simulations for the average Landauer resistance $\langle R/T \rangle$ vs $\delta$
for the same system as in Fig. \ref{avT_vs_delta}.  The main figure shows the gross-structure behavior using a semilog scale,
as well as regimes A and B.
The inset shows the fine structure for $\delta \approx \pi$ with a linear scale.  At $\delta = \pi$ is a peak, consistent with the dip observed in Fig. \ref{avT_vs_delta} for $\langle T \rangle$.  The agreement between simulation and theory is excellent.   The statistical error bar 
%for this sample and for $\delta = \pi$ is  $\sim 10^{-5}$.
is the same as in Fig. \ref{avT_vs_delta}.
}
}
\label{av_resist_vs_delta}
\end{figure}

\begin{figure}[ht]
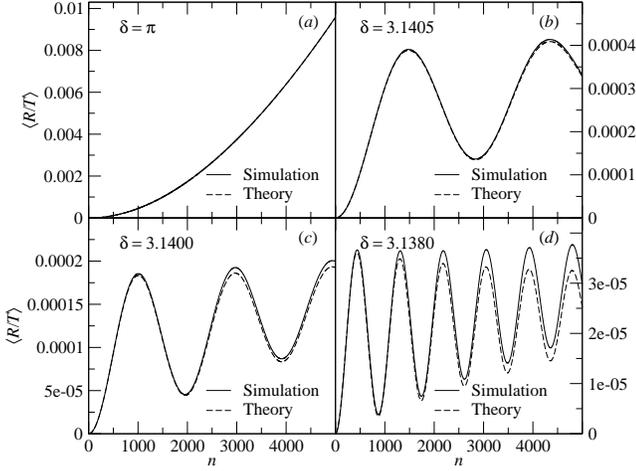

\onefigure[scale=0.38]{Fig4.eps}
\caption{
\footnotesize{
Simulations for $\langle R/T \rangle$ vs $n$, using an ensemble of $10^6$ realizations, with $y_0=0.09$, and 
for four values of $\delta \approx \pi$.
(a) $\delta = \pi$.
(b-d) $\delta = 3.1405$, $3.1400$, $3.1380$
(the results are symmetric around $\delta = \pi$ in the vicinity of this value).
The statistical error bar is in all cases smaller than $10^{-6}$ and is not indicated.
Also shown is the analytical solution of the differential 
Eq. (\ref{dA/dn,db/dn}) for the four values of $\delta$.
}
}
\label{av_resist_vs_n_delta_near_pi}
\end{figure}

In the study of the model described above we find two distinct regimes of behavior. 
In what we shall call regime A, the {\em gross-structure} behavior of the average transmission coefficient $\langle T \rangle$ as a function of $\delta$ and for a fixed number of scatterers $n$ shows ``bumps" near $\delta=\pi, \; 2\pi, \;\cdots$, and, correspondingly, the gross structure of the average Landauer resistance \cite{landauer} 
$\langle R/T \rangle$ 
($R$= reflection coefficient; $T$= transmission coefficient) shows ``valleys". 
This is illustrated in Fig. \ref{avT_vs_delta} for 
$\langle T \rangle$ and in 
Fig. \ref{av_resist_vs_delta} for 
$\langle R/T \rangle$, which include the region $\delta \sim \pi$, this being the only multiple of $\pi$ that we shall consider in what follows.  
For the case of weak scatterers, the system is almost transparent in regime A, and regime B (shown in Figs. \ref{avT_vs_delta} and 
\ref{av_resist_vs_delta}) is more localized.
This
%%%%%%%%%%%%%%%%%
%%%%%%%%%%%%%%%%%%%
%%%%%%%%%%%%%%%%%%%%%%%%%%%%%%%%%%%%%%
%%%%%%%%%%%%%%%%%%%%%%%%%%%%%%%%
gross-structure behavior is not entirely surprising. The transmission coefficient $T$ for a single barrier with fixed width and strength
becomes completely transparent ($T=1$) at the resonance values  
${\bar k}l_c= n\pi$, $n=1,2,\cdots$, where $\bar{k}$ is the wave number in the region of the barrier ($\delta \gtrsim \pi$ for low barriers).  
For a well, $T=1$ at $\delta \lesssim \pi$.  
For fixed step width and random strength with zero average, 
$\langle T \rangle$ reaches a maximum value smaller than unity at $\delta=\pi$. As the number of scatterers $n$ increases, the ``giant resonance" seen in the gross structure of $\langle T \rangle$ as a function of $\delta$ is still similar to the above description for one random scatterer.  
However, for large $n$ the system response becomes richer for values of $\delta$ very close to $\pi$ (this is a small subregion of regime A, to be denoted by $\delta \approx \pi$), where it shows a remarkable 
{\em fine structure}. 
This is illustrated in Figs. \ref{avT_vs_delta} and \ref{av_resist_vs_delta}.  
This fine structure is enhanced exactly at $\delta = \pi$, where $\langle T \rangle$ shows a dip (Fig. \ref{avT_vs_delta}) and $\langle R/T \rangle$ a maximum (Fig. \ref{av_resist_vs_delta});
i.e., the trend of the system to delocalize as it approaches $\delta = \pi$ from both sides reverses in an extremely narrow window around $\delta = \pi$, where the system is less delocalized.  The behavior of $\langle R/T \rangle$ can also be analyzed as a function of $n$ for fixed $\delta$.  For regime B, 
$1 \lesssim \delta \lesssim 2.9$ and $\delta > 3.4$, the familiar exponential increase of $\langle R/T \rangle$ with $n$ is found;the mean-free-path (mfp) increases as regime A is approached, although this trend is reversed in the neighborhood of $\delta = \pi$, consistent with the above description in relation with Figs. \ref{avT_vs_delta} and \ref{av_resist_vs_delta}.  The behavior of $\langle R/T \rangle$ as a function of $n$ is shown for four values of $\delta \approx \pi$ 
in Fig. \ref{av_resist_vs_n_delta_near_pi}.
In this regime a small change in $\delta$ changes drastically the behavior of $\langle R/T \rangle$ as a function of $n$:
from a monotonic increase ($\delta = \pi$) to an oscillating one 
($\delta \approx \pi$), the wavelength decreasing away from $\delta = \pi$.
%%%%%%%%%%%%%%%%%%%%%%%%%%%%%%%%%%%%%%
%%%%%%%%%%%%%%%%%%%%%%%%%%%%%%%%%%%%%%

It is important to mention that gross and fine structures have been seen in the response of a variety of physical systems, and have been associated with various physical mechanisms.  One of the first examples came from nuclear physics, where the structures were related to the many-body nature of the problem and gave rise to different time scales in the response~\cite{feshbach}.  
In the present case, although a very precise analytical description of the numerically observed 
i) gross structure, 
ii) fine structure and 
iii) enhanced behavior of the latter at $\delta = \pi$ 
is given ahead, an intuitive, physical explanation of ii) and iii) remains to be found.  
Indeed, Figs.~\ref{avT_vs_delta}-\ref{av_resist_vs_n_delta_near_pi} show the comparison of simulations with the theory.  
For the average resistance, the theoretical results are based on Eqs. (\ref{Land-res}), (\ref{beta.beta*-BB m=1,tilde_l 0}) and the analytical solution of (\ref{dA/dn,db/dn}); for the average transmission coefficient, on Eq. (\ref{<T>_Melnikov}) and the explanation around Eq. (\ref{T(lambda)regimeA}), in regimes B and A, respectively. 
The agreement between the two is generally found to be excellent.  
This understanding and the extremely sensitive response suggest that experimental realizations of the model could lead to important devices.

%%%%%%%%%%%%%%%%%%%%%%%%
\section{The theoretical model}
\label{theory}

The theoretical treatment follows a familiar path.  For the $r$-th scatterer (see Fig.~\ref{rand_steps}), define the dimensionless parameter $y_r = U_r l_c^2$ as a measure of its strength, where $U_r=2mV_r/\hbar^2$.  Since $y_r=(U_r/k^2)\delta^2$ 
(recall that $\delta = k l_c$), for a repulsive barrier, $U_r>0$, $\delta > \sqrt{y_r}$  means  impinging above the top of the barrier.  
The transfer matrix for the $r$-th scatterer shown in Fig. \ref{rand_steps} 
has the structure 
\cite{mello-kumar}
%%%%%%%%%%%%%%%%
\begin{equation}
M_r=
\left[
\begin{array}{cc}
\alpha_r & \beta_r \\
\beta_r^* & \alpha_r^*
\end{array}
\right] ,
\end{equation}
%%%%%%%%%%%%%
due to time-reversal invariance.
One finds 
%%%%%%%%%%%%%%%%
\begin{subequations}
\begin{eqnarray}
\alpha_r
&=& e^{-i \delta} \left[\cos\left(\sqrt{\delta^2-y_r} \; \right) \right.
\nonumber \\
&&
%\hspace{1cm}
+ \left. i \; \frac{2\delta^2-y_r}{2 \delta \sqrt{\delta^2-y_r}}
\sin\left(\sqrt{\delta^2-y_r}\right)\right],
\label{alphar} \\
\beta_r
&=& -i e^{-i (2r-1) \delta} \frac{y_r}{2 \delta \sqrt{\delta^2-y_r}} 
\sin (\sqrt{\delta^2-y_r}).
\label{betar}
\end{eqnarray}
\label{Mr}
\end{subequations}
%%%%%%%%%%%%%%%%
Landauer's resistance for this scatterer is $R_r/T_r=|\beta_r|^2$. 

%%%%%%%%%%%%%%%%%%%%%%%
\subsection{Landauer resistance of the chain}
\label{landauer res. chain}

Consider next a chain of $n$ scatterers to which one more, to be called a {\em building block} (BB), is added.  
Combining the corresponding transfer matrices, the average resistance for the chain, $\langle |\beta^{(n)}|^2 \rangle$, obeys a recursion relation that couples with the quantity $\langle \alpha^{(n)}\beta^{(n)} \rangle$ as
%%%%%%%%%%%%%%%%
\begin{subequations}
\begin{eqnarray}
& A(n+1)-A(n)
= 2\langle |\beta_{n+1}|^2 \rangle A(n)
\nonumber \\
&
\hspace{5mm} 
+ 2\left[\langle \alpha_{n+1}  \beta^{*}_{n+1}\rangle 
\langle \beta^{(n)} \alpha^{(n)} \rangle  + {\rm c.c.} \right] ,
\label{recursion_M,m=1 a}  \\
&\langle \alpha^{(n+1)}\beta^{(n+1)} \rangle 
- \langle \alpha^{(n)}\beta^{(n)} \rangle 
= \langle \alpha_{n+1}  \beta_{n+1}\rangle A(n)
\nonumber \\
&
+\left(\langle \alpha^2_{n+1}\rangle -1 \right) 
\langle \alpha^{(n)}\beta^{(n)} \rangle
+\langle   \beta^2_{n+1} \rangle \langle \alpha^{(n)}\beta^{(n)} \rangle^* \; .
\label{recursion_M,m=1 b}
\end{eqnarray}
\label{recursion_M,m=1}
\end{subequations}
%%%%%%%%%%%%%%%%
We have defined 
%%%%%%%%%%%%%%%%%%%%%%%%
\begin{equation}
A(n)= 1+2 \langle |\beta^{(n)}|^2 \rangle,
\label{A(n)}
\end{equation}
%%%%%%%%%%%%%%%%%%%%%%%%
where $\langle \cdots \rangle$ indicates an ensemble average and ${\rm c.c.}$ denotes the complex conjugate.  
The nature of the ensemble follows from the definition of the statistical model: each $y_r$ is uniformly distributed over the interval 
$(-y_0, y_0)$ (then
$
\langle y_r  \rangle = 0
\label{<y_r>}
$,
$
\langle y_r^2  \rangle = y_0^2 /3
$)
and different $y_r$'s are statistically independent.  
The recursion relations, Eq.~(\ref{recursion_M,m=1}), are exact and thus take into account {\em all multiple scattering processes} occurring in the chain. 

A comment on the scaling property of the model parameters is in order.
Although the transfer matrix of a single scatterer $r$ depends, in principle, on the 3 parameters $E$, $U_r$ and $l_c$,
Eq. (\ref{Mr}) shows that these 3 parameters occur in the combinations $\delta$ and $y_r$. 
Thus, for a specific realization of disorder, the resistance 
$(R/T)^{(n)}$ of an $n$-step chain is a function of the $n+2$ parameters $\delta, n, y_1, \cdots y_n$.
An average over realizations, i.e., over $y_1, \cdots y_n$, is performed with the distribution function just described, which depends on the parameter $y_0$ only.
Thus the resulting average resistance $\langle (R/T)^{(n)} \rangle$
depends only upon the combination of 3 parameters: $\delta$, $n$ and 
$y_0\equiv U_0/l_c^2$, instead of 4. 
In Figs. \ref{avT_vs_delta} and \ref{av_resist_vs_delta}, $y_0$ and $n$ are kept fixed and $\delta$ is varied. This could be realized, e. g., by using a fixed $l_c$
throughout the chain of fixed length $n$ and varying the energy.
In each of the panels in Fig. \ref{av_resist_vs_n_delta_near_pi}, $y_0$ and $\delta$ are kept fixed and $n$ is varied. This could be realized, e. g., by fixing $l_c$ and the energy, and varying the length of the chain $n$.

Equation (\ref{recursion_M,m=1}) admits two distinct approximations, which aid analytical treatment and correspond to regimes B and A, respectively.  Define
%%%%%%%%%%%%%%%%
\begin{equation}
K_1=\frac
{\left|\langle \alpha_{n+1}  \beta^{*}_{n+1}\rangle \;
\langle \beta^{(n)} \alpha^{(n)} \rangle  + {\rm c.c.} \right|}
{\langle |\beta_{n+1}|^2 \rangle \; \left[1+2\langle |\beta^{(n)}|^2 \rangle \right]} \; ,
\label{K}
\end{equation}
%%%%%%%%%%%%%%%%
which is the ratio of the ``coupling" containing
$\langle \beta^{(n)} \alpha^{(n)} \rangle$
to the ``direct part" containing
$\langle |\beta^{(n)}|^2 \rangle$) on the right-hand side of Eq. (\ref{recursion_M,m=1 a}).
Numerically, it appears that the coupling can be neglected in regime B: for $n \lesssim 5000$ (larger values of $n$ have not been checked
for the analysis of $K_1$) 
and for the value $y_0 = 0.09$ (corresponding to weak scatterers) to be used throughout this paper 
(except in Fig. \ref{av_resist_vs_delta_y0=0.01}), 
$K_1 \lesssim 1 \%$.  Dropping the coupling terms on the RHS of  Eq. (\ref{recursion_M,m=1 a}) gives
%%%%%%%%%%%%%%%%
\begin{equation}
A(n+1)-A(n)
=  2\langle |\beta_{1}|^2 \rangle \; A(n) \; ,
\label{recursion_A(n) cut}
\end{equation}
%%%%%%%%%%%%%%%%
since 
$\langle |\beta_{n+1}|^2 \rangle = \langle |\beta_{1}|^2 \rangle$.
The solution of the recursion relation (\ref{recursion_A(n) cut}) with the initial condition $A(0)=1$ is
%%%%%%%%%%%%%%%%
%%%%%%%%%%%%%%%%
%\begin{subequations}
\begin{equation}
A(n)= \left( 1+2 \langle |\beta_1|^2 \rangle \right)^n 
%\label{A(n)-sol a} 
= e^{2 n \frac12 \ln \left(1+2 \langle |\beta_1|^2 \rangle \right)} 
%\label{A(n)-sol b}
\equiv e^{2 n l_c / \ell} .
%\label{defn lc/lhat}
\label{A(n)-sol}
\end{equation}
%\end{subequations}
%%%%%%%%%%%%%%%%
This is the well known exponential increase found by Landauer \cite{landauer}, with
%%%%%%%%%%%%%%%%
\begin{equation}
\frac{l_c}{\ell}
= \frac12 \ln \left(1+2 \langle |\beta_1|^2 \rangle \right) \; .
\label{eff_loc_length}
\end{equation}
%%%%%%%%%%%%%%%%
For $\langle |\beta_1|^2 \rangle = \langle R_1/T_1 \rangle  \ll 1$, 
$1 / \ell$ is approximately the average reflection coefficient per unit length, which is identified with the {\em inverse mean free path} (mfp) \cite{froufe_et_al_2007} and, in the present 1D problem, is of the order of the {\em inverse localization length}.

Landauer's average resistance for the chain is thus
%%%%%%%%%%%%%%%%
\begin{equation}
\langle |\beta^{(n)}|^2 \rangle
= \frac12 ( e^{2 n l_c / \ell} - 1).
\label{Land-res}
\end{equation}
%%%%%%%%%%%%%%%%
Using Eq. (\ref{betar}), the quantity $\langle |\beta_1|^2 \rangle$ 
%appearing in Eqs. (\ref{A(n)-sol}) and (\ref{eff_loc_length}) 
is
%%%%%%%%%%%%%%%%
\begin{equation}
\langle |\beta_1|^2 \rangle
= \left\langle   
\frac{y_1^2}{4 \delta^2 (\delta^2-y_1)} 
\sin ^2\left( \sqrt{\delta^2-y_1}\right)  
\right\rangle \; ,
\label{exact <beta1^2>}
\end{equation}
%%%%%%%%%%%%%%%%
where the ensemble average has to be taken with the distribution of $y_1$ described above.
It can be evaluated exactly, but for small values of $y_0/\delta^2$ we find 
%%%%%%%%%%%%%%%%
%\begin{subequations}
\begin{equation}
\langle |\beta_{1}|^2 \rangle
= \frac{l_c}{\tilde{\ell}} + O\left(\frac{y_0}{\delta ^2}\right)^4 ,
%\label{beta.beta*-BB m=1}
\hspace{5mm}
\frac{l_c}{\tilde{\ell}}
= \frac{y_0^2}{12} \; \frac{\sin^2 \delta}{\delta^4} \; ,
%\label{tilde_l 0}
\label{beta.beta*-BB m=1,tilde_l 0}
\end{equation}
%\end{subequations}
%%%%%%%%%%%%%%%%%%%%%%%%%%%%%%%
where the approximate expression $\tilde{\ell}$ for the mfp shows explicitly its dependence on the strength $y_0$ and the spatial extension $l_c$ of the steps.  For the value $y_0=0.09$ used here, the approximate expression (\ref{beta.beta*-BB m=1,tilde_l 0}) is an excellent approximation to the exact result.

The result of Eq.~(\ref{Land-res}), with the approximate mfp given in Eq.~(\ref{beta.beta*-BB m=1,tilde_l 0}), was used to plot the theoretical average Landauer resistance in Fig.~\ref{av_resist_vs_delta} for 
$1\lesssim \delta \lesssim 2.9$ and $\delta > 3.4$ (regime B) and is seen to give an excellent description of the data.  Indeed, the results of the numerical simulations literally lie on top of the theoretical ones.  
Therefore regime B is marked by the lack of importance of the coupling terms
(our first approximation).

It is not possible to treat the behavior in regime A
($2.9 \lesssim \delta \lesssim 3.4$) without accounting for the coupling terms in Eq. (\ref{recursion_M,m=1}).
In particular, well inside regime A ($\delta \approx \pi$) one finds
considerable fine structure (see Fig.~\ref{av_resist_vs_n_delta_near_pi}) which cannot be explained without the coupling.
Figure \ref{av_resist_vs_n_delta_near_pi}a shows the average Landauer resistance precisely for $\delta = \pi$ as a function of the number of scatterers $n$, obtained from a simulation.
The approximate inverse mfp gives $l_c/\tilde{\ell}=0$ for $\delta = \pi$, and the theoretical result of Eq. (\ref{Land-res}) with $\ell$ replaced by $\tilde{\ell}$ would give $\langle |\beta^{(n)}|^2 \rangle \equiv 0$, in gross contradiction with the increasing behavior with $n$ obtained numerically.  The inverse mfp of Eq. (\ref{eff_loc_length}) using the exact average, Eq.~(\ref{exact <beta1^2>}), is non-zero for $\delta=\pi$ (it is found to be of $O(10^{-9})$), but not large enough to explain the numerical data.  
Indeed, for $\delta$ in the edges of regime A, the coupling terms in Eq. (\ref{recursion_M,m=1 a}) become important and, for 
$\delta \approx \pi$, dominate and the uncoupled difference equation 
(\ref{recursion_A(n) cut}) becomes a bad approximation
(also, it does not reproduce the result shown in the inset in Fig. \ref{av_resist_vs_delta}).  

This motivates the second approximation, which is a continuum approximation that converts the recursion relations, 
Eq. (\ref{recursion_M,m=1}), into differential equations.  However, to do this the ensemble averages $\langle |\beta^{(n)}|^2 \rangle$ and $\langle \beta^{(n)} \alpha^{(n)} \rangle$ must behave ``smoothly" as functions of $n$.  For this purpose consider the recursion relations similar to those of Eq.~(\ref{recursion_M,m=1}), except with a BB containing $m$ scatterers.  It is their behavior as a function of $m$ which is of interest.  Evaluating analytically the various BB quantities appearing in these equations in the same approximation as 
Eq. (\ref{beta.beta*-BB m=1,tilde_l 0}), i.e.~neglecting terms of order 
$(m y_0/\delta^2)^4$, leads some of them to have the form 
$g(\delta)(e^{2im\delta}-1)$.
For example, we find
%%%%%%%%%%%%%%%%
\begin{subequations}
\begin{align}
%\begin{eqnarray}
\langle \alpha_{n+1,n+m} \; \beta_{n+1,n+m}^* \rangle
&=\frac{m l_c}{\tilde{\ell}} f_1(\delta,m) e^{2in\delta} 
%\nonumber \\
%&& \hspace{3.5cm} 
+ O\left(\frac{m y_0}{\delta^2}\right)^4 , 
\nonumber
\\
\label{alpha.beta*-BB 0}
\\
f_1(\delta,m)
&= i \delta \; \frac{\sin \delta - \delta e^{i \delta}}{\sin^3 \delta} \; 
\frac{e^{2im\delta}-1}{2im\delta} \; .
\label{f1 0}
\end{align}
%\end{eqnarray}
\label{<lambda>,<ab>smooth 0}
\end{subequations}
%\end{align}
%%%%%%%%%%%%%%%%
This result 
%of (\ref{alpha.beta*-BB 0}) 
is, in general, not linear in $m$ and thus
{\em prevents us from making a continuous approximation
to convert the difference equations into differential equations}.
However, for $\delta=\pi$ these BB expressions either vanish or are proportional to $m$, and for $\delta \sim \pi$ (regime A) they are all proportional to $m$.  In these cases we have also verified numerically that 
$\langle |\beta^{(n+1)}|^2 \rangle 
\approx \langle |\beta^{(n)}|^2 \rangle $
and $\langle \beta^{(n+1)} \alpha^{(n+1)} \rangle
\approx \langle \beta^{(n)} \alpha^{(n)} \rangle$,
i.e., that these quantities behave smoothly as functions of $n$, when $n$ is changed by one unit.
Thus, in this regime it is justified to make a continuous approximation, which gives the coupled differential equations
%%%%%%%%%%%%%%%%%%%%%%%%%%%%%%%
\begin{subequations}
\begin{eqnarray}
&&\frac{\partial A(n)}{\partial n}
= 2 \; \frac{l_c}{\tilde{\ell}} \;
\Big[ A(n) + f_1(\delta) b(n) +  f_1^*(\delta) b^*(n) \Big]
\label{dA/dn}
\\
&& \frac{\partial b(n)}{\partial n}
= - \frac{l_c}{\tilde{\ell}} f_4(\delta) e^{2i \delta} A(n)
\nonumber \\
&&
+\left[
(e^{2i \delta} -1) - \frac{l_c}{\tilde{\ell}} f_2(\delta) e^{2i \delta}
\right]b(n)
%\nonumber \\
%&&
-\frac{l_c}{\tilde{\ell}} f_3(\delta)e^{2i \delta}  b^*(n) \; ,
\nonumber \\
\label{db/dn}
\end{eqnarray}
\label{dA/dn,db/dn}
\end{subequations}
%%%%%%%%%%%%%%%%%%%%%%%%%%%%%%%
neglecting terms $O(y_0/\delta^2)^4$.
Here, $f_1(\delta)$ is obtained from the above $f_1(\delta,m)$ setting $m=1$, and $f_2(\delta), f_3(\delta), f_4(\delta)$ are similar functions of $\delta$.
The quantity $A(n)$ is defined in Eq. (\ref{A(n)}) and
%%%%%%%%%%%%%%%%%%%%%%%%%%%%%%%
\begin{equation}
b(n)
= e^{2in\delta} \langle \alpha^{(n)}\beta^{(n)} \rangle.
\label{b(n)}
\end{equation}
%%%%%%%%%%%%%%%%%%%%%%%%%%%%%%%
Equation (\ref{dA/dn,db/dn}) is subject to the initial conditions
%%%%%%%%%%%%%%%%%%%%%%%%%%%%%%%
\begin{equation}
A(0)=1, \;\;\;  b(0) = 0 .
\label{in. conds.}
\end{equation}
%%%%%%%%%%%%%%%%%%%%%%%%%%%%%%%
The explicit expression for 
$f_2(\delta), f_3(\delta), f_4(\delta)$ and the exact analytical solution of Eqs. (\ref{dA/dn,db/dn}) will be given elsewhere.

That exact solution was used to plot the theoretical average resistance as a function of $\delta$ in Fig. \ref{av_resist_vs_delta}, in regime A
($2.9 < \delta  < 3.4$). 
Since the coupling has been taken fully into account, the agreement between theory and simulations is excellent.  
Thus  Eq. (\ref{dA/dn,db/dn}) gives a theoretical quantitative description of the fine structure, including the prominent peak at $\delta=\pi$ shown in the figure.

The analytical solution of Eq. (\ref{dA/dn,db/dn}) was also used to plot $\langle R/T \rangle$ as a function of $n$ in 
Fig. \ref{av_resist_vs_n_delta_near_pi}a-d.  The agreement with the simulations is excellent exactly for $\delta=\pi$;
the remarkable oscillatory behavior for $\delta \approx \pi$ is a result of the coupling, and is generally reproduced quite well by the theory;  
the agreement deteriorates further away from $\delta = \pi$ 
(see also right inset in Fig. \ref{avT_vs_delta}).
There is evidence that the deterioration starts for larger $n$'s as $y_0$ decreases.  
Agreement is also excellent for $\Re \langle \alpha \beta \rangle^{(n)}$ and $\Im \langle \alpha \beta \rangle^{(n)}$ for $1 \leq n \leq 5000$ and for the same values of $\delta$ shown in Fig. \ref{av_resist_vs_n_delta_near_pi} (these results will be presented elsewhere).

%%%%%%%%%%%%%%%%%%%%%%%%
\subsection{Transmission coefficient of the chain}
\label{T for chain}

Finally, we analyze the transmission coefficient behavior for the chains studied above.  It is convenient to use the polar representation of Ref.~\cite{mello-kumar} and introduce the notation $\lambda_r = |\beta_r|^2$ for the $r$-th scatterer and 
$\lambda^{(n)}=|\beta^{(n)}|^2$ for the chain consisting of $n$-scatterers.  In this notation, Eq. (\ref{recursion_A(n) cut}) is written as
[see Eq. (\ref{beta.beta*-BB m=1,tilde_l 0})]
%%%%%%%%%%%%%%%%%
\begin{subequations}
\begin{eqnarray}
\frac{\Delta \langle \lambda^{(n)} \rangle}{\Delta \tilde{s}}
\approx 1+2 \langle \lambda^{(n)} \rangle,
\end{eqnarray}
%%%%%%%%%%%%%%%%%
where
%%%%%%%%%%%%%%%%%
\begin{eqnarray}
\tilde{s}&=& \frac{L}{\tilde{\ell}}, \\
\Delta \tilde{s} &=& \langle \lambda_1 \rangle = \frac{l_c}{\tilde{\ell}} \; .
\end{eqnarray}
\label{delta lambda}
\end{subequations}
%%%%%%%%%%%%%%%%%%%%
Its continuous approximation is
%%%%%%%%%%%%%%%%%%%%%%
\begin{equation}
\frac{\partial \langle \lambda \rangle_{\tilde{s}}}{\partial \tilde{s}}
= 1+2 \langle \lambda \rangle_{\tilde s} \; .
\label{differential evolution equation for <lambda>}
\end{equation}
%%%%%%%%%%%%%%%%%%%%%%%
The evolution of $\langle \lambda \rangle_{\tilde{s}}$ 
coincides with that found from the evolution equation for the $\lambda$-probability density, $w_{\tilde{s}}(\lambda)$,  
known as Melnikov's equation \cite{mello-kumar}
%%%%%%%%%%%
\begin{equation}
\frac{\partial w_{\tilde{s}}(\lambda)}{\partial {\tilde s}}
= \frac{\partial}{\partial \lambda} \left[\lambda(1+\lambda)
\frac{\partial w_{\tilde{s}}(\lambda)}{\partial \lambda}\right].
\label{melnikov}
\end{equation}
%%%%%%%%%%%%%%%%%%%%%%%
A similar result holds true for the second moment of $\lambda$; a quantity $K_2$ can be defined, analogous to $K_1$ defined in
Eq.~(\ref{K}), which is also found numerically to be small ($< 1.6 \%$) in regime B.
It is not possible to verify numerically the values of the analogous quantities $K_p$
(defined for the $p$-th moment) for all moments;
but in fact it is not possible even for an individual moment if the order $p$ is too large, because numerical control is lost in the calculation due to the rapid exponential increase of these moments with $\tilde{s}$.  
We thus propose the approximate validity of Melnikov's equation for regime B; this assumption allows finding the statistical properties of $T$ which, in terms of $\lambda$, can be written as
%%%%%%%%%%%
\begin{equation}
T=\frac{1}{1+\lambda} \; .
\label{T(lambda)}
\end{equation}
%%%%%%%%%%%%%%%%%%%%%%%
From Melnikov's Eq. (\ref{melnikov}), the expression for the $p$-th moment of the transmission coefficient can be reduced to quadratures, with the result
\cite{mello_1991}
%%%%%%%%%%%
\begin{equation}
\langle T^p \rangle
= \frac{2 {\rm e}^{-s/4}}{\Gamma(p)}
\int_0^{\infty}{\rm e}^{-st^2}
\left|\Gamma(p-\frac12 +it)\right|^2  t \; \tanh (\pi t) {\rm d} t,
\label{<Tp>_Melnikov}
\end{equation}
%%%%%%%%%%%
which, for the first moment, gives
%%%%%%%%%%%
\begin{equation}
\langle T \rangle
= 2 {\rm e}^{-\tilde{s}/4}
\int_0^{\infty}{\rm e}^{-\tilde{s}t^2}
\pi t [\tanh (\pi t)/\cosh(\pi t)] {\rm d}t.
\label{<T>_Melnikov}
\end{equation}
%%%%%%%%%%%
This result was compared with numerical experiments in Fig. \ref{avT_vs_delta}, 
for $2.5 < \delta < 2.9$ and 
$3.4 < \delta < 4$ (regime B).  The agreement is excellent.

In regime A, the theoretical analysis uses
%%%%%%%%%%%
\begin{equation}
\langle T \rangle {\approx 1 - \langle \lambda \rangle},
\label{T(lambda)regimeA}
\end{equation}
%%%%%%%%%%%
since 
$\langle \lambda \rangle \ll 1$ (Fig. \ref{av_resist_vs_delta}), and 
$\langle \lambda \rangle$ is taken from the analytical solution of Eq. (\ref{dA/dn,db/dn}).  
This approximation for $\langle T \rangle$ was compared with the numerical simulations in Fig. \ref{avT_vs_delta}. The agreement is excellent.
The pronounced dip observed for $\langle T \rangle$ at $\delta=\pi$ is consistent with the peak observed in Fig. \ref{av_resist_vs_delta} for the average resistance.

To complement the above analysis, we may consider chains in which the potential strengths $V_r$ are not distributed symmetrically around zero, and thus have a non-zero average.
When all the $V_r > 0$, one expects the ``bump" in $\langle T \rangle$ of 
Fig. \ref{avT_vs_delta} to move to $\delta > \pi$, and, when all the $V_r < 0$, to $\delta < \pi$.
This behavior is indeed found in numerical simulations, which also show a similar shift in the position of the dip, along with a loss of the symmetry around this shifted position.

\begin{figure}[ht]
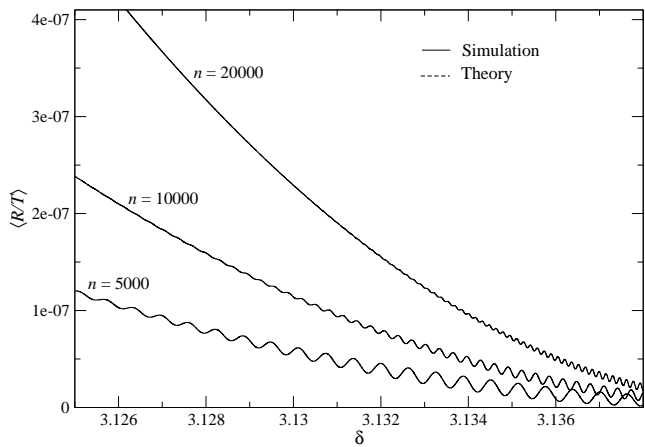

\onefigure[scale=0.35]{Fig5.eps}
\caption{
\footnotesize{
Theory and simulations for the average Landauer resistance 
$\langle R/T \rangle$ vs $\delta$
for three chains, with $n=5000$, $10000$ and $20000$ scatterers, respectively, and $10^5$ realizations.
The decrease, with increasing $n$, of the wavelength and amplitude of the oscillations and the region where they are conspicuous, is clearly seen.
The value $y_0=0.01$ was chosen to show these features more clearly.
The agreement between theory and simulations is excellent.
The statistical error bar is $< 10^{-9}$ and is not indicated.
}
}
\label{av_resist_vs_delta_y0=0.01}
\end{figure}

Finally, a few comments are in order with regards to the asymptotic behavior of our theory for large $n$'s and the validity of the theoretical analysis. 

a) If the calculations of Figs. \ref{avT_vs_delta} and \ref{av_resist_vs_delta} are repeated for various $n$'s: 
i) The ``bump" in $\langle T \rangle$ and the ``valley" in $\langle R/T \rangle$ become narrower as $n$ increases from $n=1000$ to $n=5000$;
it has been found, albeit with a smaller number of realizations (1000), that this tendency persists to $n=50000$. 
ii) The maximum in the bump in $\langle T \rangle$ remains undisturbed.
iii) The peak in the average resistance for $\delta = \pi$ becomes ever higher as $n$ is increased, in full accordance with the behavior shown in Fig. \ref{av_resist_vs_n_delta_near_pi}a,
while the dip in the average transmission for $\delta = \pi$ becomes ever deeper as the chain is made longer.
Effects i), ii), iii) are reproduced well by the theory.

b) As for the fine structure oscillations seen in Fig. 
%\ref{avT_vs_delta} and 
\ref{av_resist_vs_delta} for $\langle R/T \rangle$ 
{\em as function of $\delta$} for a fixed number $n=5000$ of scatterers, the calculations have been extended to $n=20000$. 
As $n$ is increased, various features are observed
(zooming in on the results, to make them visible)
in the region away from $\delta=\pi$ where one can define a wavelength
and an amplitude of the oscillations (for $n=5000$ this region would be outside the extremely narrow window shown in the inset in Fig. 3):
i) the wavelength decreases;
ii) the amplitude decreases;
iii) the extension of the region away from $\delta=\pi$ where the oscillations are present before damping out, decreases.
An example of these three features is shown in 
Fig. \ref{av_resist_vs_delta_y0=0.01}, where an excellent agreement between theory and simulations is also seen.

%%%%%%%%%%%%%%%%%%%%%%%%%%%%%%%%%%%%%%
%%%%%%%%%%%%%%%%%%%%%%%%%%%%%%%%

c) The fact, apparent from 
Fig. \ref{av_resist_vs_n_delta_near_pi}, that, 
{\em as function of $n$} (for fixed $\delta$), the fine structure oscillations damp out as $n$ increases, has been seen to persist to $n=100000$ 
and is reproduced well by the theory.

d) It is to be remarked that our theoretical analysis is restricted to small values of $y_0$,
as was indicated right after Eq. (\ref{dA/dn,db/dn}).
Therefore, although one expects, for a fixed $y_0$, theory and numerical simulations to deviate starting from some $n_M$, 
there is evidence that $n_M$ can be made larger by decreasing $y_0$. This fact was already remarked at the end of the previous subsection.

%%%%%%%%%%%%%%%%%%%%%
\section{Conclusions}
To summarize, we have discussed the problem of wave transport in 1D disordered systems consisting of barriers and wells with a finite, constant width $l_c$, and random strength.  For weak scatterers, the system is almost transparent for $\delta=k \ell_c \sim \pi$, and less delocalized farther away.  For $\delta \approx \pi$, one observes a remarkable situation: a fine structure behavior which is enhanced exactly at $\delta = \pi$, where the system becomes {\em less delocalized}.  In this region, a small change in $\delta$ modifies drastically the behavior of the average resistance as a function of $n$.
All of these phenomena are seen in simulations and are described very well by the theoretical analysis. We want to stress that our theory is a fully analytical theory, which, in regime B, is given by decoupling Eqs. (\ref{recursion_M,m=1}), and in regime A by the coupled Eqs. (\ref{dA/dn,db/dn}) which, in turn, represent a continuum approximation to Eqs. (\ref{recursion_M,m=1}).
The theory has full predictive power, with no adjustable parameters.

The success of our theoretical analysis and the extreme sensitivity we have described suggest the importance of the system experimental realization.

%%%%%%%%%%%%%%%
\acknowledgments

M. D. acknowledges financial support by CONACyT, M\'exico, through scholarship 234662, and P. A. M. and M. Y. through grant 79501.
S. T. gratefully acknowledges financial support from the U.S.~National Science Foundation grant PHY-0855337.
The authors are grateful to J. J. S\'aenz for important discussions and suggestions.
Technical support in the computer simulations from C. L\'opez Natar\'en is gratefully acknowledged.
%%%%%%%%%%%%%%%%%%%%%%%%%

\end{document}